\documentclass{llncs}
\usepackage{llncsdoc}
\usepackage{bm}
\usepackage{graphicx}

\renewcommand{\v}[1]{\mathbf{\bm #1}}

\title{Semantic vector space models predict neural responses to complex visual stimuli}

\author{Umut G\"u\c{c}l\"u\inst{1} \and Marcel A. J. van Gerven\inst{1}}

\institute{Radboud University, Donders Institute for Brain, Cognition and Behaviour,\\P.O. Box 9104, Nijmegen, the Netherlands}

\begin{document}

\maketitle
\begin{abstract}
Encoding models have as their objective to predict neural responses to naturalistic stimuli with the aim of elucidating how sensory information is represented in the brain. This prediction is achieved by representing the stimulus in terms of a suitable feature space and using this feature space to linearly predict observed neural responses. Here, we investigate to what extent semantic vector space models can be used to predict neural responses to complex visual stimuli. We show that these models provide good predictions of neural responses in downstream visual areas, improving significantly over a low-level control model based on Gabor wavelet pyramids. The outlined approach provides a new way to model and map high-level semantic representations across cortex.
\end{abstract}

\section{Introduction}

A principal goal in sensory neuroscience is to understand how properties of our environment are reflected in neural response patterns. This can be achieved by constructing encoding models that explicitly link stimuli to responses via an intermediate feature space. In literature, several feature spaces have been proposed such as linear~\cite{Schoenmakers2013} and non-linear~\cite{Kay2008}  feature spaces. More recently, research has focused on the development of feature spaces that are themselves learnt from a large amount of training data which embodies the statistical invariances in our environment~\cite{Guclu2014a}. To date, the best predictions of neural responses to naturalistic stimuli have been achieved using deep neural networks~\cite{Guclu2015}. An alternative strategy for extracting relevant features from a complex naturalistic stimulus is to directly annotate stimuli in terms of their semantics. This approach has been used by Huth et al.~\cite{Huth2012}, who hand-labelled individual movie frames according to the entities that make up the WordNet ontology~\cite{Miller1995}. It was demonstrated that these entities can be used to predict neural responses across cortex. This is close in spirit to the work by Mitchell et al.~\cite{Mitchell2008}, who showed that neural responses to presented nouns could be predicted using a feature space that couples nouns to their associated verbs.

The goal of the present work is to take the sketched approach to the next level by using semantic vector space models of language that represent individual words in terms of continuous vector representations to predict neural responses to naturalistic stimuli~\cite{Mikolov2013}. These word embeddings have proven highly successful in the sense that the estimated continuous vector representation can be shown to accurately capture semantic information. If we  are able to use these word embeddings to predict neural responses to naturalistic stimuli then this may provide new insights into the neural representation of semantic information. Moreover, it may offer an elegant way to reconstruct semantic information from patterns of brain activity.

Here, we tested this approach by annotating individual images with semantic labels (i.e. words) and using the associated word embeddings to predict neural responses as people experienced those stimuli. We show that neural responses in downstream areas are accurately predicted by the word embeddings, demonstrating for the first time that the semantic space implied by word embeddings provide insights in the neural representation of complex naturalistic stimuli.

\section{Methods}

\subsection{Experimental Data}

In order to examine the suitability of word embeddings for predicting neural responses, we made use of the data set that was originally published in~\cite{Kay2008}. 

For each of two male subjects (S1 and S2), five sessions of data were collected as subjects were presented with natural images. Training and test data were collected in the same scan sessions. The total number of images used for training and testing were 1750 and 120, respectively. Each training image was repeated two times, and each test image was repeated 13 times. In this study, we analyzed the data from one subject (S1).

Stimuli consisted of grayscale natural images drawn randomly from different photographic collections. Subjects fixated on a central white square. Stimuli were flashed at 200 ms intervals for 1 s followed by 3 s of gray background.

Data were acquired using a 4 T INOVA MR scanner and a quadrature transmit/receive surface coil. Eighteen coronal slices were acquired covering occipital cortex (slice thickness 2.25 mm, slice gap 0.25 mm, field of view 128$\times$128 mm$^2$). fMRI data were acquired using a gradient-echo EPI pulse sequence (matrix size 64$\times$64, TR 1 s, TE 28 ms, flip angle 20$^\circ$, spatial resolution 2$\times$2$\times$2.5 mm$^3$). 

fMRI scans were coregistered and used to estimate voxel-specific response timecourses. After deconvolution of these timecourses from the time series data, an estimate of response amplitude was obtained for each presented unique image in each voxel. Voxels were assigned to visual areas using retinotopic mapping data acquired in separate sessions. Additionally, anatomical and functional volumes were coregistered manually. \textit{FreeSurfer}\footnote{http://surfer.nmr.mgh.harvard.edu/} and \textit{MrTools}\footnote{http://gru.stanford.edu/doku.php/mrtools/overview} were used for cortical surface reconstruction and visualization, respectively.

\subsection{Construction of feature spaces}

In order to associate word embeddings with naturalistic image stimuli, the following procedure was used. First, each of the total of 1870 images was manually annotated by an annotator who labeled each image with the first word that came to mind when inspecting the contents of those images. This led to a total of 512 unique labels for the stimuli in the training set and 81 unique labels for the stimuli in the test set (73 overlapping labels between the training and test set). Next, each word was converted to a continuous vector representation. 

To obtain continuous word embeddings we used different models based on the skip-gram word-to-vec (W2V) model described in~\cite{Mikolov2013a} which learns word representations by predicting the surrounding word of a given word and the global vectors for word representations (GLoVe) model described in~\cite{Pennington2014} which learns word representations by factorizing a word co-occurance matrix. For the W2V model, we used the pretrained 300-dimensional word representations trained on 100 billion work Google News dataset provided online\footnote{https://code.google.com/p/word2vec}. For the GloVe model, we used eight pretrained word representations trained on 6 billion word Wikipedia 2014 + Gigaword 5 (50-, 100-, 200- and 300-dimensional) and 27 billion word Twitter (25-, 50-, 100-, 200-dimensional) datasets provided online\footnote{http://nlp.stanford.edu/projects/glove}. For a stimulus $\v{x}$, we use $\textrm{vec}(w(\v{x}))$ to denote the word embedding associated with the label $w(\v{x})$ assigned to that stimulus. 

As a control model, to compare low-level feature spaces with high-level feature spaces, we used the Gabor wavelet pyramid (GWP) model described in~\cite{Kay2008}.

\subsection{Predicting neural responses}

The continuous word embeddings were used to predict neural responses to individual stimuli by using them as input to a linear response model. A separate response model was trained for each voxel using regularized linear regression. The used estimation procedure is described in detail in~\cite{Guclu2014a}. For each voxel $i$, after estimating the regression coefficients $\v{\beta}_i$, we obtain $\v{\mu}_i (\v{x})=\v{\beta}_i^T \textrm{vec}(w(\v{x}))$ as the predicted response of that voxel to input stimulus $\v{x}$. Voxel response models were estimated using the entire training set and evaluated on the test set. A five-fold cross-validation on the training set was used for model selection (regularization parameters and significant voxels). We discarded the voxels whose cross-validated prediction accuracies on the training set were not significantly higher than chance level ($p < 0.05$, Student's t-test, Bonferroni correction). 

To quantify how well the used feature space predicts voxel responses, we defined the encoding performance of a model as the Pearson correlation coefficient ($r$) between the observed and predicted responses on the test set. We also projected the encoding performance back onto the cortical surface to get an estimate of which regions are sensitive to the semantic information that is embodied by the continuous word vector representations.

We compared the encoding performance of multiple models across the union of the significant voxels for the models being compared since different voxels can be significant for different models. We set the encoding performance for the nonsignificant voxels that are in the union to zero.

\section{Results}

\subsection{Comparison of continuous word embeddings}

We first compared the mean encoding performance of the 9 different semantic vector space models (Figure 1). Recall that the models differed in architecture (W2V or GloVe), dimensionality (25, 50, 100, 200 or 300 dimensions), training set (Google News, Wikipedia 2014 + Gigaword 5, Twitter) and training set size (100B, 6B or 27B words). The 300-dimensional W2V model trained on the 100 billion word Google News dataset had the highest mean encoding performance (0.2064), whereas the 25-dimensional GloVe model trained on the 27 billion word Twitter dataset had the lowest mean encoding performance (0.1655). The difference in architecture, dimensionality and training set size did not have as much an impact on the mean encoding performance as did the dataset selected for training the model. For example, the encoding performance of the 300-dimensional W2V model trained on the 100 billion word Google News dataset was significantly higher than that of the 50- and 100-dimensional GloVe models trained on the 27 billion word Twitter dataset but not the 6 billion word Wikipedia 2014 + Gigaword 5 dataset.

\begin{figure}[ht!]
\centering
\includegraphics{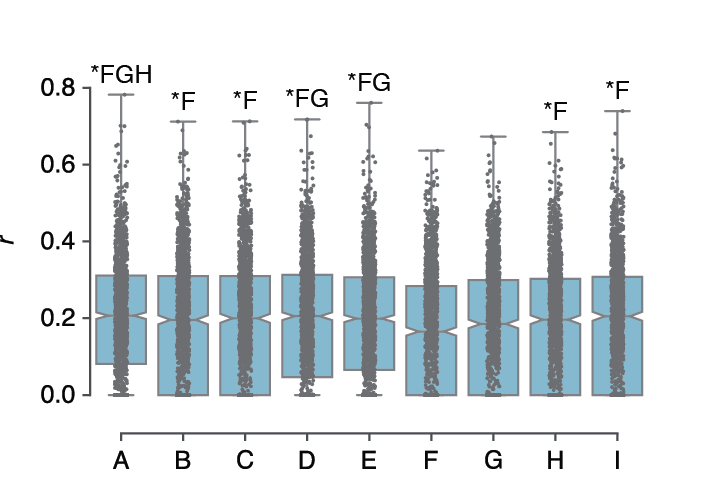}
\caption{Comparison of the encoding performance of the 9 different semantic vector space models (A: W2V 300D; Google News 100B. B-E: GloVe 50D, 100D, 200D and 300D; Wikipedia 2014 + Gigaword 5 6B. F-I: GloVe 25D, 50D, 100D and 200D; Twitter 27B). The letters above the box plots indicate significantly different mean encoding performance ($p < 0.05$, one-way analysis of variance, Tukey's range test). For example, the mean encoding performance of model A is significantly higher than that of models F, G and H but not models B, C, D, E and I.}
\end{figure}

\subsection{Cortical mapping of semantic information}

We then analyzed how the results of the best performing model (W2V model) change across the voxels in V1, V2, V3, V3A, V3B, V4, LO and anterior to V1-LO. The fraction of the significant voxels for the W2V model and the mean encoding performance of the W2V model significantly increased from upstream to downstream areas. The lowest and highest fraction of significant voxels were in V3 (3.4\%) and LO (19.1\%), respectively. The lowest and highest mean encoding performance were in V2 (0.1679) and LO (0.2672), respectively. However, 54.9\% of the significant voxels were anterior to V1-LO. The mean prediction accuracy for these voxels was 0.2657.

\begin{figure}[ht!]
\centering
\includegraphics{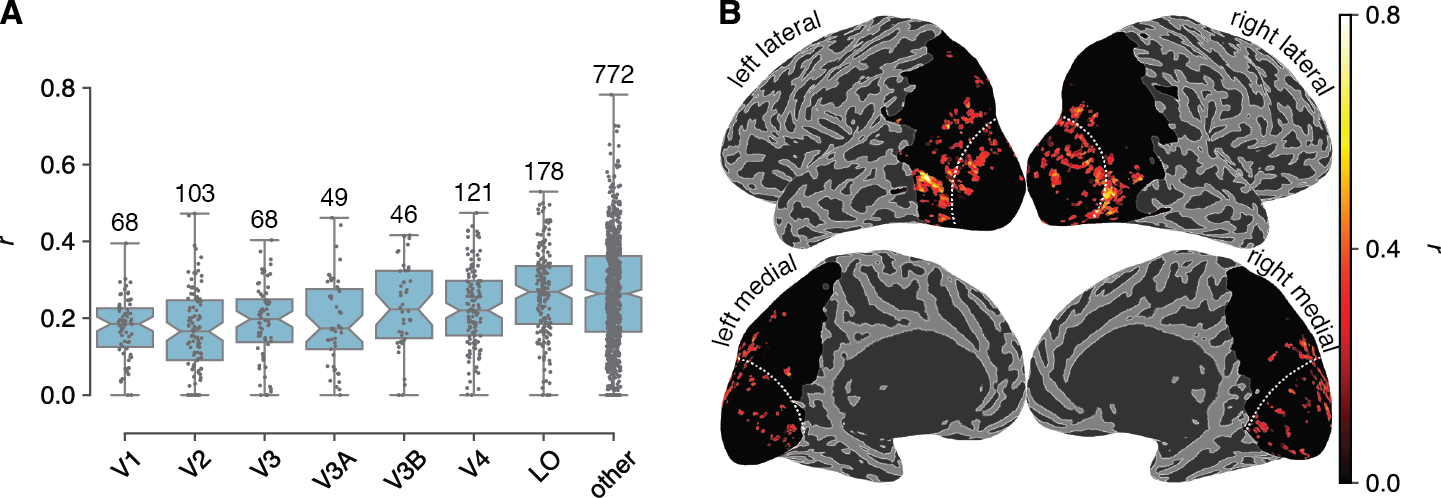}
\caption{Results of the W2V model. (A) Encoding performance of the W2V model across the voxels in V1, V2, V3, V3A, V3B, V4, LO and anterior to V1-LO. The number of significant voxels for the W2V model are shown above the box plots. (B) Projection of the encoding performance of the W2V model to the cortical surface. The dashed curve separates the voxels in V1-LO (posterior) from the rest of the voxels (anterior).}
\end{figure}

\subsection{Comparison with control models}

To validate our findings, we compared the mean encoding performance of the W2V model with that of the GWP model (Figure 3). Recall that the models differ in selectivity to level of stimulus features (low- or high-level). The difference in mean encoding performance of the models decreased from low- to mid-level areas (V1, V2, V3, V4) and increased from mid- to high-level areas (V3A, V3B, LO and anterior to V1-LO). The W2V model outperformed the GWP model in high- but not low- and mid-level areas. In each of the areas, the mean encoding performance of one of the models was significantly higher than that of the other model ($p < 0.05$, $p < 0.001$ or $p \ll 0.001$, two-sample Student's t-test).

\begin{figure}[ht!]
\centering
\includegraphics{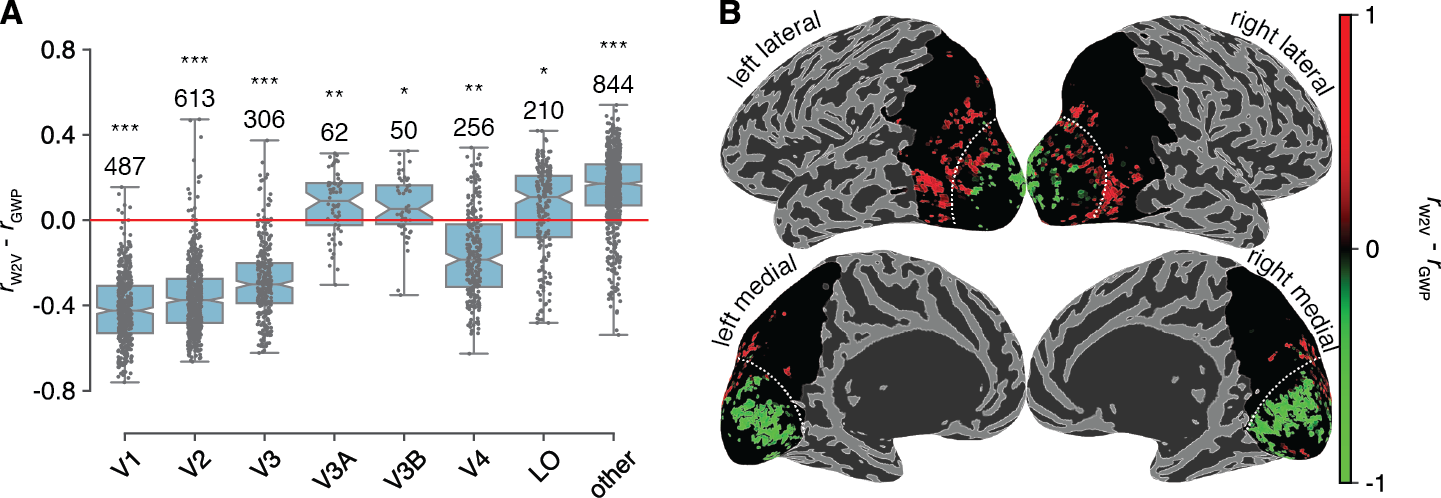}
\caption{Comparison of the encoding performance of the W2V model with that of the GWP model. (A) Encoding performance of the W2V vs GWP models across the voxels in V1, V2, V3, V3A, V3B, V4, LO and anterior to V1-LO. The number of significant voxels for both of the models are shown above the box plots. Asterisks above the box plots indicate significantly different mean encoding performance (*: $p < 0.05$, **: $p < 0.001$ and ***: $p \ll 0.001$, two-sample Student's t-test. (B) Projection of the encoding performance of the W2V vs GWP models to the cortical surface. The dashed curve separates the voxels in V1-LO (posterior) from the rest of the voxels (anterior).}
\end{figure}

\section{Conclusion}

Semantic vector space models that learn a projection of words to a vector space that preserves the meaning of words, similarities between words and analogies have been very successful in computational linguistics. However, there has been no study on the relationship between such word embeddings and human brain activity.

Here, we investigated this relationship by testing the extend to which learned embeddings of visual stimulus labels in a vector space are predictive of human brain activity. We showed that stimulus-driven voxel responses in downstream visual areas are accurately captured by such word embeddings. That is, we demonstrated for the first time the relationship between the semantic space implied by word embeddings and the neural representation of complex naturalistic stimuli.

Our findings are in line with the previous studies which show that individual concepts are represented in the inferior temporal cortex but expand on them by showing that a continuous, distributed, low-dimensional vector space can be underlying these representations. Future work should provide insights into how concepts are represented in the brain in the absence of complex visual stimulation (e.g. encoding visual or auditory words) or in combination with other concepts (e.g. encoding visual or auditory sentences).

We consider this research to be the next step in the quest to elucidate how brains encode their environment. Furthermore, it could one day provide the basis for exciting new applications in brain-computer interface technology.

\section{Acknowledgements}

The data used in this paper is archived at CRCNS.org under digital object identifier http://dx.doi.org/10.6080/K0QN64NG. This research was supported by VIDI grant number 639.072.513 of The Netherlands Organization for Scientific Research (NWO).

\newcommand{\etalchar}[1]{$^{#1}$}

\end{document}